\documentclass{svproc}
\usepackage{url}
\usepackage{natbib}
\makeatletter
\def\@biblabel#1{}
\makeatother

\begin{document}
\newcommand{\aap}{{Astron. Astrophys.}}
\newcommand{\aaps}{{Astron. Astrophys. Suppl.}}
\newcommand{\apj}{{Ap. J.}}
\newcommand{\apss}{{Astrophys. Space Sci.}}
\newcommand{\grl}{{Geophys. Res. Lett.}}
\newcommand{\solphys}{{Solar Phys.}}
\newcommand{\apjl}{{Ap. J. Lett.}}
\newcommand{\apjs}{{Ap. J. Suppl.}}
\newcommand{\araa}{{Annu. Rev. Astron. Astrophys.}}
\newcommand{\arnp}{{Annu. Rev. Nucl. Part. Sci.}}
\newcommand{\nat}{{Nature}}
\newcommand{\mnras}{{MNRAS}}
\newcommand{\prd}{{Phys. Rev. D}}
\newcommand{\prc}{{Phys. Rev. C}}
\newcommand{\physrep}{{Phys. Rep.}}
\newcommand{\pasa}{{Publ. Astron. Soc. Austr.}}
\newcommand{\pasj}{{Publ. Astron. Soc. Jap,}}
\newcommand{\nar}{{New Astron. Rev.}}
\newcommand{\npa}{{Nucl. Phys. A}}
\newcommand{\jgr}{{J. Geophys. Res.}}
\newcommand{\memsai}{{Mem.~Soc.~Astron.~Italiana}}
\newcommand{\qjras}{{Quart. J. Royal Astron. Soc.}}
\newcommand{\ssr}{{Space Sci. Rev.}}
\newcommand{\RMP}{{Rev. Mod. Phys.}}
\mainmatter              
\title{Explosive Nucleosynthesis: What we learned and what we still do not understand}
\titlerunning{Explosive Nucleosynthesis}  
%
\author{Friedrich-Karl Thielemann\inst{1,2}}
\authorrunning{F-K. Thielemann} 
%
%
\institute{University of Basel, CH-4056 Basel, Switzerland,\\
\email{f-k.thielemann@unibas.ch},\\ 
\and
GSI Helmholtz Center for Heavy Ion Research,\\
D-64291 Darmstadt, Germany}

\maketitle              

\begin{abstract}
This review touches on historical aspects, going back to the early days of nuclear astrophysics, initiated by B$^2$FH and Cameron, discusses (i) the required nuclear input from reaction rates and decay properties up to the nuclear equation of state, continues (ii) with the tools to perform nucleosynthesis calculations and (iii) early parametrized nucleosynthesis studies, before (iv) reliable stellar models became available for the late stages of stellar evolution. It  passes then through (v) explosive environments from core-collapse supernovae to explosive events in binary systems (including type Ia supernovae and compact binary mergers), and finally (vi) discusses the role of all these nucleosynthesis production sites in the evolution of galaxies. The focus is put on the comparison of early ideas and present, very recent, understanding.
\end{abstract}
\section{Introduction}
In this short text it is not possible to give an overview over more than 60 years of nuclear astrophysics, and especially explosive nucleosynthesis, but we try to address the questions how it all started, what nuclear and technical input was/is required, how intitially parametrized calculations developed into full scale (magne-to-)hydrodynamic simulations, which nucleosynthesis processes take place, and how they impact ejecta compositions, which again make their way into galactic evolution and the present solar abundance composition. The solar neutrino problem is solved, the expansion of the Universe understood with the aid of type Ia superovae, the role of neutron star mergers was clarified in 2017 with GW170817. Many other open questions remain, especially how the abundance evolution from lowest metallicities to present can be understood.

\section{Nuclear Input and Reaction Networks}

\subsection{Nuclear Input}

Over the years an enormous wealth of experimental information entered into compilations of nuclear reaction rates, initially with
the lead of the Kellogg group around Willy Fowler (Fowler, Caughlan, Zimmerman 1967, Fowler, Caughlan, Zimmerman 1975,
and Caughlan \& Fowler 1988). Major updates were due to Angulo et al. (1999, NACRE), Xu et al. (2013, NACRE II), followed 
by Iliadis et al. (2001), Longland et al. (2010) I, Iliadis et al. (2010abc) II, III, IV from North Carolina, which entered into Starlib (Sallaska et al. 2013). Uncountable individual investigations have been undertaken by many groups, here especially the Wiescher group should be mentioned, clarifiying recently the role of $^{12}$C($\alpha,\gamma)^{16}$O in stellar helium burning (deBoer et al. 2017). And the LUNA Lab at Gran Sasso plays a key role to measure minute reaction cross sections at lowest energies deep underground, important in early hydrostatic stellar burning stages (see https://luna.lngs.infn.it/index.php/scientific-output/publications). An enormous body of neutron capture reactions has been established by Bao \& K\"appeler (1987), and Bao et al. (2000), expanded more recently by the KADONIS collaboration (Dillmann et al. 2014). This has been extended to unstable nuclei by the n\_ToF collaboration at CERN. A constantly growing set of results, involving highly unstable nuclei, comes from radioactive ion beam facilities at MSU, GSI, RIKEN, GANIL, and Lanzhou. 

Theoretical developments to predict nuclear reaction rates started with Truran et al. (1966), utilizing the statistical (Hauser-Feshbach) model with ground state properties, continuing with Truran (1968), Michaud \& Fowler (1970, 1972, adding improved optical potentials), Arnould (1972, the first one including excited states), Arnould \& Beelen (1974), and Truran (1972). Holmes, Woosley, Fowler, Zimmerman (1976), and Woosley, Fowler, Holmes, Zimmerman (1978) made use of level densities via a back-shifted Fermi gas. Thielemann, Arnould, Truran (1986, 1988) developed the SMOKER code, Rauscher \& Thielemann (2000, 2004) extended it to the NoSMOKER approach. Panov et al. (2010) included fission. Present efforts center around T. Rauscher, extending NoSMOKER to SMARAGD (Rauscher 2011), and S. Goriely et al., who started out with the MOST code in 1997 (Goriely 1997),  introducing since then many improvements with applications of the TALYS code (e.g. Goriely et al. 2008).
    
Weak interaction rates, like e.g. beta-decays, electron captures, neutrino interactions are of equal importance, pioneered by Kratz et al.
(1986),  M\"oller \& Kratz (1997) to measure/predict beta-decay half-lives; Fuller, Fowler, Newman (1985), Langanke \& Martinez-Pinedo (2003), addressing electron capture rates as well as neutrino interactions with nuclei. Many others followed, like e.g. Marketin et al. (2016), focussing on neutron-rich nuclei far from stability, especially important for the r-process. All predictions of such efforts for unstable nuclei, addressing also fission, 
have an initimate relation to properties of nuclear mass models (see e.g. Sobiczewski et al.  2018).
Combined information has entered Complete Reaction Libraries, e.g. presently publicly available Reaclib, Bruslib, Starlib, as well as
the Equation of State database CompOSE  (Oertel et al. 2017).

\subsection{Reaction Neworks}

Early approaches to solve nuclear reaction networks, which are stiff systems of ordinary differential equations and not solvable with the means of explicit methods, were undertaken by Truran et al. (1966, 1967) and Arnett \& Truran (1969). The solution via the 
implicit backward Euler method was obtained in a linear approach. Woosley et al. (1973),  Arnould (1976), Thielemann et al. (1979),  changed this to a fully converged multi-dimensional Newton-Raphson scheme. Restricted nuclear networks have long been
used in stellar evolution codes (e.g. Iben 1985).  Presently in use on a global basis are BasNet (going back to Thielemann et al. 1979), NetGen (in Bruslib),  XNET (Hix \& Thielemann 1999), Timmes et al. (1999), Cabezon et al. (2004), NucNet (Meyer \& Adams 2007), 
WinNet (Winteler et al. 2012), SkyNet (Lippuner \& Roberts 2017).

\section{Stellar Models}

First explosive nucleosynthesis calculations were all based on parameter studies rather than realistic stellar models, but were highly important to explore results. All explosive burning stages, from H, He, C, Ne, O, Si-burning to nuclear statistical equilibrium (NSE) have been tested by these early investigations
e.g. expl. Si-burning:  Fowler \& Hoyle (1964),  Bodansky et al. (1968), expl. O- and Si-burning: Woosley, Arnett, Clayton (1973), expl. Ne- and C-burning: Arnett (1969a), Howard et al. (1972), Truran and Cameron (1978), Arnett \& Wefel (1978), Morgan (1980).
As initial stellar models examined only early burning stages, these investigations had to be done via parameter studies with
assumed adiabatic expansions from initial peak temperatures and densities. First attempts to model late burning stages and provide pre-collapse models for supernova explosions were undertaken by Arnett (1977), Weaver, Woosley, Zimmerman (1978, leading to the Kepler code), and Nomoto \& Hashimoto (1988).
Presently highly sophisticated input exists from Chieffi \& Limongi (2018, FRANEC), Heger \& Woosley (2010, KEPLER), Meynet, Hirschi and collaborators (e.g. Georgy et al. 2013, GENEC), Paxton et al. (2011, MESA), Umeda/Yoshida (e.g. Yoshida et al. 2016), Nakamura et al. (2015). Stellar models have been verified by the solution of the solar neutrino problem (McDonald, 2016).

\section{Type Ia Supernovae}
Binary systems with accretion onto one compact object can lead to (depending on the accretion rate) explosive events with thermonuclear runaway (under electron-degenerate conditions). In case of accreting white dwarfs  this
can cause nova or type Ia supernova explosions. The explanation of type Ia supernovae goes back to Hoyle \& Fowler (1960). First carbon-detonation models were developed by Arnett (1969b), Arnett et al. (1971), and Woosley (1986), later discarded as they did not fit observations.  Iben \& Tutukov (1984) and Webbink (1984) laid the theoretical groundwork for so-called single and double degenerate systems, depending whether one white dwarf is or two white dwarfs are involved in the binary system. First 1D deflagration models were developed by Nomoto et al. (1982ab, 1984) and Woosley \& Weaver (1986). M\"uller \& Arnett (1986) and later
Khoklov, M\"uller \& H\"oflich (1993) started general combustion approaches. Consistent ignition modeling for degenerate condition is approached with the MAESTRO code (Zingale et al. 2011). Presently, single-degenerate systems starting with central carbon deflagration, double degenerate mergers, He-accretion caused double detonations, and even white dwarf collisions are considered (for a review see e.g. Thielemann et al. 2018). Major progress is due to observations, disentangling the possible scenarios (Maoz et al. 2014, Noebauer et al. 2017, Goldstein \& Kasen 2018).
Important understanding for the combination of contributing scenarios comes from their nucleosynthesis of Mn ($^{55}$Co-decay) and Zn in galactic evolution (Seitenzahl \& Townsley 2017, H\"oflich et al. 2017, Leung \& Nomoto 2018, Mishenina et al. 2015,Tsujimoto \& Nishimura 2018).

\section{Core-Collapse Supernovae (CCSNe)}
We are on the path of solving the core-collapse supernova problem in a self-consistent way. While early approaches assumed that the bounce of the collapsing Fe-core at nuclear densities would permit a sufficiently energetic shock front and an explosion, this has been shifted to explosions driven by neutrinos (Bethe 1990). There exists a growing set of 2D and 3D CCSN explosion simulations  (see e.g. reviews by Janka et al.  2012, 2016, Burrows 2013, 2018, Bruenn et al. 2016, Foglizzo et al. 2015, Nakamura et al. 2015, Cabezon et al. 2018). Active groups are based in Garching/Belfast/Monash/RIKEN (Janka, M\"uller, M\"uller, Just..), Princeton/Caltech/MSU (Burrows, Ott, Couch..), Oak Ridge (Mezzacappa, Hix, Lenz, Messer, Harris ..), Tokyo/Kyushu (Takiwaki, Nakamura, Kotake), Paris (Foglizzo et al.), and Basel (Liebend\"orfer, Cabezon, Hempel ...). Open questions relate to the stellar mass limit where core-collapse ends in black hole formation (Pan et al. 2018, Kuroda et al. 2018), and when - due to rotation and magnetic fields - this leads to hypernovae (Nomoto et al. 2013).
 
To provide complete nucleosynthesis predictions from self-consistent multi-D simulations is still in its infancy.
For this reason 1D approximations, based on piston or thermal bomb approaches have been undertaken for many years (e.g. Thielemann et al. 1996, Heger \& Woosley 2010, Nomoto et al. 2013, Limongi \& Chieffi 2018). They lack self-consistent predictions of explosion energies, mass cuts between neutron star and ejecta, as well as the neutron-richness ($Y_e$) 
of the innermost ejecta. More recently improved 1D approximations have followed, attempting to mimic multi-D effects and avoiding the shortfalls mentioned above
(Ugliano et al. 2012, Perego et al. 2015, Ertl et al. 2016, Sukhbold et al. 2016, Ebingeret al. 2018).
A major role in determining the composition of the innermost ejecta play neutrino interactions with outflowing matter. Opposite to early ideas that matter close to the proto-neutron star would be neutron-rich, neutrino capture on neutrons (favored by the neutron-proton mass difference) is winning against antineutrino capture on protons and turns matter (slightly) proton-rich, causing a 
$\nu$p-process (Fr\"ohlich et al. 2006ab, Pruet et al. 2006, Wanajo 2006, Eichler et al. 2018, Curtis et al. 2018). While such a $\nu$p-process is supported by present simulations, an r-process is apparently not occurring, at most a weak r-process
(Martinez-Pinedo et al. 2012, Roberts et al. 2012, Arcones \& Thielemann 2013). For possible exceptions, in case of fast rotation and strong magnetic fields, see the next section.

\section{Origin of the Heavy Elements}
The production of a fraction of the heavy elements up to Pb and Bi has long been postulated since B$^2$FH (1957) and Cameron (1957) via the slow neutron capture (s-) process in shell He-burning during stellar evolution (see e.g. K\"appeler et al. 2011). 
The origin of the heaviest elements up to Th, U, and Pu (and beyond) via the rapid neutron capture (r-) process remained a puzzle until very recently, despite much progress in understanding the nuclear physics impact (see e.g. Cowan et al. 1991, Kratz et al. 1993, Arnould et al. 2007, Petermann et al. 2012, Goriely \& Martinez-Pinedo 2015). Regular core-collapse supernovae were champions for many years (see early ideas in Cowan et al. 1991 and neutrino-wind powered models in Woosley et al. 1994, Takahashi et al. 1994 or later in Farouqi et al. 2010), but  apparently they cannot provide the conditions required (Freiburghaus et al. 1999a), as matter turns rather proton- than neutron-rich via neutrino interactions (see previous section). A rare fraction of magneto-rotational supernovae, dependent on initial rotation rates and magnetic fields, seems to have a chance for succeeding (Cameron 2003, Nishimura et al. 2006, Fujimoto et al. 2008, Winteler et al. 2012, Nishimura et al. 2015, 2017, M\"osta et al. 2015, 2018, Halevi \& M\"osta 2018). Neutron star mergers have been proven to support the conditions for a full r-process since GW170817. For a review before this event, from the early proposals (Lattimer \& Schramm 1974, Eichler et al. 1989), over first simulations (Davies et al. 1994, Ruffert \& Janka 1997, Rosswog et al. 1999) and the first nucleosynthesis predictions (Freiburghaus et al. 1999b) up to early 2017 see Thielemann et al. (2017). Numerous investigations have followed this observational break-through by e.g. Barnes, Hotokezaka, Kasen, Metzger, Rosswog, Tanaka, Wollager (for references see Horowitz et al. 2018).

\section{Chemical Evolution and Explosive Nucleosynthesis}
Since B$^2$FH (1957) and Cameron (1957) we know that essentially all elements beyond H and He are made in stars. The Big Bang produced only H, He, and some Li (Cyburt et al. 2017). The production of heavier elements as a function of time/metallicity depends on occurrence frequencies and delay times for individual nucleosynthesis contributions in galaxies, scrutinized by ever improving observational facilities (Li et al. 2018).
The observed enhanced value of  [$\alpha$/Fe] abundance ratios ($\alpha$=O, Ne, Mg, Si, S, Ca, Ar, Ti) at low metallicities, turning down to solar values at metallicities from [Fe/H]=-1 to 0, due to the input of Ni and Fe-enhanced type Ia supernovae, is reasonably well understood
since Matteucci \& Greggio (1986), Wheeler et al. (1989), Timmes et al. (1995), Kobayashi et al. (2006), Matteucci (2012), and Nomoto et al. (2013).
Recent supernova models (Curtis et al. 2018) can explain the Fe-group. A more interesting question is which role Mn and Zn play in this game. The low value of [Mn/Fe] at low metallicities, rising above [Fe/H]=-1, indicates its origin as $^{55}$Co from type Ia supernovae (Mishenina et al. 2015).
[Zn/Fe]=0, i.e. solar values, can be made in regular core-collapse supernovae (Fr\"ohlich et al. 2006a, Curtis et al. 2018), but the upturn (with a sizable scatter)  below [Fe/H]=-2 has to be due to hypernovae and/or a certain class of magneto-rotational supernovae 
(Tsujimoto \& Nishimura 2018). The reason that [Zn/Fe] stays at 0 also beyond [Fe/H]=-1, when type Ia supernovae take over the production of Fe-group elements, indicates that there must also exist a type Ia subclass producing Zn (as $^{64}$Ge), caused probably by He-detonations, permitting a strong alpha-rich freeze-out.

The large scatter of [Eu/Fe] at low metallicities by more than two orders of magnitude (Sneden et al. 2008, Roederer et al. 2014, Hansen et al. 2018) indicates a rare site for the strong r-process. This could be consistent with neutron star mergers, but also be due to (still only proposed) magneto-rotational supernovae (their existence being supported by the observations of magnetars as endpoints of such events, Greiner et al. 2015). Chemodynamical galactic evolution calculations have been performed e.g. by Argast et al. (2004), Cescutti et al. (2015), van de Voort et al. (2015), Shen et al. (2015), Wehmeyer et al. (2015), Hirai et al. 2017, Cot\'e et al. (2018), Hotokezaka et al. (2018). There exists sufficient supporting material that neutron star mergers are probably the main contributor for the solar r-process composition, but they occur with a delay in galactic evolution which causes problems explaining the [Eu/Fe] ratios at metallicites as low as  [Fe/H]=-3. So-called actinide boost stars, i.e. objects found at such low metallicities with enhanced Th/Eu and U/Eu ratios, have probably not the typical solar r-process origin. Their features seem explainable by an interplay between the r-process path and fission properties.  Conditions which are slightly less neutron-rich in magneto-rotational supernovae than in neutron star mergers possibly support such resulting final compositions (Holmbeck 2018ab). Similar results are found by Eichler and Wu (private communication).

{\bf Notes and Comments:}
I want to thank the organizers of NIC XV at Gran Sasso for asking me to present this special invited talk. It gave me challenges to cover an extended research field from its beginnings up to present. I enjoyed especially following the ongoing progress and thank M. Busso and G. Meynet for advice.
I apologize that this review is probably biased and also not complete. It omits almost completely nucleosynthesis in stellar evolution and explosive events in novae, X-ray bursts and superbursts.
It would not have been possible without the insight I obtained working with my collaborators and students.
Thanks go to my teachers (Arnett, Arnould, Cameron, Fowler, Hilf, Hillebrandt, Schramm, Truran), my long-term 
collaborators outside Basel (Cowan, Kratz, Langanke, Nomoto, Panov, Wiescher), all PhD students, often supervised jointly within the extended Basel group (Brachwitz, Dillmann, Ebinger, Eichler, Fehlmann, Freiburghaus, Frensel, Fr\"ohlich, Heinimann, Hix, K\"appeli, Liebend\"orfer, Mocelj, Oechslin, 
Perego, Reichert, Rembges, Rosswog, Scheidegger, Wehmeyer), and my present/former Basel co-workers (Arcones, Cabezon, Hempel, Hirschi, Kolbe, Kuroda, Liebend\"orfer, Martinez-Pinedo, Nishimura, Pignatari, Pan, Rauscher), of whom many have dispersed around the world. 
\vspace{-10cm}%
%

\end{document}